\documentclass[prb,showpacs,twocolumn,amsmath,amssymb]{revtex4}

\usepackage{bm}
\usepackage{graphicx}
\usepackage{bbold}

\newcommand{\expectv}[1]{\langle #1 \rangle}

\newcommand{\pdagger}{{\phantom{\dagger}}}
\newcommand{\dt}{\Delta\tau}

\newcommand{\reff}[1]{Fig.\ \ref{fig:#1}}

\usepackage{color}

\definecolor{darkgreen}{rgb}{0,0.7,0}

\begin{document}

\title{Orbital-selective Mott transitions in the anisotropic\\
two-band Hubbard model at finite temperatures}

\author{C.~Knecht}
\author{N.~Bl\"umer}
\email{Nils.Bluemer@uni-mainz.de}
\author{P.\ G.\ J.\ van Dongen}
\affiliation{Institute of Physics, Johannes Gutenberg University, 55099 Mainz, Germany}

\date{\today}

  \begin{abstract}
    The anisotropic degenerate two-orbital Hubbard model is studied
    within dynamical mean-field theory at low temperatures.
    High-precision calculations on the basis of a refined quantum
    Monte Carlo (QMC) method reveal that two distinct orbital-selective Mott
    transitions occur for a bandwidth ratio of 2 even in the absence
    of spin-flip contributions to the Hund exchange. The second
    transition -- not seen in earlier studies using QMC, iterative
    perturbation theory, and exact diagonalization -- is clearly
    exposed in a low-frequency analysis of the self-energy and in
    local spectra.
  \end{abstract}
  \pacs{71.30.+h, 71.10.Fd, 71.27.+a}
  \maketitle


  The Mott-Hubbard metal-insulator transition -- a nonperturbative
  correlation phenomenon -- has been a subject of fundamental interest
  in solid state theory for decades.\cite{Gebhard97} Recently, this
  field became even more exciting by the discovery \cite{Nakatsuji00a,
    Nakatsuji00b} of a two-step metal-insulator transition in the
  effective 3-band system Ca$_{2-x}$Sr$_{x}$RuO$_4$, for which the
  name \emph{orbital-selective} Mott-transition (OSMT) was coined.
  \cite{Anisimov02} The Ca$_{2-x}$Sr$_{x}$RuO$_4$ system was
  investigated theoretically in detail by Anisimov {\em et al.}\cite{Anisimov02} 
  within the local density approximation (LDA and
  LDA+U) and within dynamical mean-field theory\cite{Georges96} (DMFT)
  solved using the non-crossing approximation (NCA). The underlying
  assumption of a correlation (rather than lattice-distortion) induced
  OSMT found support in further band structure calculations
  \cite{Fang01,Fang04} and strong-coupling expansions for the
  localized electrons in the orbital-selective Mott
  phase.\cite{Sigrist04}

  Microscopic studies of the OSMT usually consider the 2-band Hubbard
  model $H=H_1+H_2$, where
  \begin{eqnarray*}
  H_1&=&-\sum_{\langle ij\rangle m\sigma}  t^\pdagger_m c^{\dag}_{im\sigma}
  c^\pdagger_{jm\sigma}\,+\,U\sum_{im}n_{im\uparrow} n_{im\downarrow}\\
    &&+\sum\nolimits_{i\sigma\sigma'}(U'-\delta^\pdagger_{\sigma \sigma'} J_z)n^\pdagger_{i1\sigma}
  n^\pdagger_{i2\sigma'}
  \end{eqnarray*}
  includes hopping between nearest-neighbor sites $i,j$ with amplitude
  $t_m$ for orbital $m\!\in\!\{1,2\}$, \emph{intra}- and
  \emph{inter}\-orbital Coulomb repulsion parametrized by $U$ and
  $U'$, respectively, and Ising-type Hund's exchange coupling;
  $n^\pdagger_{im\sigma} =c^{\dag}_{im\sigma} c^\pdagger_{im\sigma}$ for spin
  $\sigma\in\{\uparrow,\downarrow\}$.  In addition,
\[
  H_2=\tfrac{1}{2}J_\perp\sum\nolimits_{im\sigma}c^\dag_{im\sigma}\left(c^\dag_{i\bar{m}\bar{\sigma}}c^\pdagger_{im\bar{\sigma}}
    +c^\dag_{im\bar{\sigma}}
    c^\pdagger_{i\bar{m}\bar{\sigma}}\right)c^\pdagger_{i\bar{m}\sigma}
  \]
  contains spin-flip and pair-hopping terms (with $\bar{1}\equiv 2$,
  $\bar{\uparrow}\equiv \downarrow$ etc.).  In cubic lattices, the
  Hamiltonian is invariant under spin rotation, $J_z=J_\perp\equiv J$;
  furthermore $U'=U-2J$. In the following, we refer to $H_1+H_2$ in
  this spin-isotropic case as the $J$-model and to the simplified
  Hamiltonian $H_1$ as the $J_z$-model.
  
  Liebsch\cite{Liebsch03a, Liebsch03b, Liebsch04} questioned the OSMT
  scenario for Ca$_{2-x}$Sr$_{x}$RuO$_4$ on the basis of
  finite-temperature quantum Monte Carlo (QMC) calculations (within
  DMFT) for the $J_z$-model using $J_z=U/4$, $U'=U/2$, and
  semi-elliptic ``Bethe'' densities of states with a bandwidth ratio
  $W_2/W_1=2$.  Additional studies using iterative perturbation
  theory (IPT)\cite{Liebsch04} seemed\cite{fn:IPT} to confirm his
  conclusion of a single Mott transition of both bands at the same
  critical $U$-value.  Meanwhile, Koga {\em et al.}\ found an OSMT using
  exact diagonalization (ED), applied to the full
  $J$-model,\cite{Koga04a} but not for the $J_z$-model.\cite{Koga04b}
  Consequently, the OSMT scenario was attributed to spin-flip and
  pair-hopping processes.

  Very recently, four preprints appeared, \cite{Ferrero05, deMedici05,
    Arita05, Koga05} in which the OSMT was investigated in detail
  within the DMFT framework. Ref.\ \onlinecite{Ferrero05} applied the
  Gutzwiller variational approach and ED to the $J$-model at
  temperature $T=0$ and confirmed the existence of an OSMT, provided
  that the ratio $W_2/W_1$ of the two band widths is sufficiently
  small. Interestingly, Ref.\ \onlinecite{Ferrero05} also suggested
  the existence of small spectral weight near the Fermi level of the
  narrow-band subsystem in the orbital-selective Mott phase. Similar
  results were obtained by de' Medici {\em et al.},\cite{deMedici05} who
  used slave-spin mean field theory (which is closely related to the
  Gutzwiller method).  Arita and Held \cite{Arita05} used the
  projective QMC method to investigate the
  $J$-model at $T=0$ and demonstrated a first OSMT for $J=U/4$ and
  $U=2.6$ (in units of half the narrow-band width). Finally, Koga et
  al.\cite{Koga04b} used QMC to characterize the OSMT for the $J$-model at
  finite $T$ on the basis of spin, charge and orbital susceptibilities
  as well as spectral functions; they further
  showed that additional hybridization between the bands smears out
  the OSMT at $T=0$. Remarkably, the problem originally investigated
  by Liebsch, \cite{Liebsch03a, Liebsch03b, Liebsch04} i.\,e., the
  $J_z$-model with $U'=U-2J_z$ was addressed only in Ref.\ 
  \onlinecite{deMedici05}, where de' Medici {\em et al.}\  found an OSMT at
  $T=0$ within slave-spin mean-field theory, in contradiction to
  Liebsch's and Koga's earlier results. Since the slave-spin method is
  essentially uncontrolled, the existence of an OSMT in the
  $J_z$-model therefore remains unclear.
  
  The goal of this work is to clarify this issue and to establish
  whether $H_1$ can be regarded as a \emph{minimal model} for the
  OSMT.  In the following, we will sketch our high-precision DMFT
  algorithm, which supplements QMC by a high-frequency expansion of
  the self-energy, before we discuss relevant observables and
  data-analysis techniques and present numerical results.

  {\it Quantum Monte Carlo scheme --} Within the DMFT, the lattice
  problem is mapped onto a single-impurity Anderson model with general
  hybridization, which has to be determined
  self-consistently.\cite{Georges96} DMFT algorithms are characterized
  by the numerical method employed in the solution of the impurity
  problem and by the iterative procedure used in order to establish
  self-consistency.  In this work, the imaginary-time Green function
  $G(\tau)$ of the multi-band impurity is obtained in discretized form from
  Hirsch-Fye QMC simulations.  Since the lattice Dyson equation is
  most easily
  formulated in the frequency domain, Fourier transformations (FTs)
  are needed twice per DMFT self-consistency cycle. 
  Since naive FTs would violate the analytic properties
  of Green functions and self-energies, all DMFT-QMC codes
  use either a special transformation\cite{Ulmke95} or interpolate the
  discrete QMC data by cubic splines.\cite{Jarrell93,Georges96} Recently, it was
  realized\cite{Bluemer02,Oudovenko02} that natural boundary
  conditions are not well suited for modeling imaginary-time Green
  functions and that  misfits can be reduced by allowing for
  non-zero second derivatives of $G$ at the boundaries $\tau\to 0$,
  $\tau\to \beta$.\cite{Oudovenko02} However, this improvement cannot eliminate
  misfits arising from higher order derivatives of $G$, which may be
  large at $0$, $\beta$ and are neglected in cubic splines.
  
  Our DMFT-QMC method further improves on this situation by applying the FTs
  to a difference Green function $\Delta G\equiv
  G_{\text{QMC}}-G_{\text{HFE}}$, where the model Green function $G_{\text{HFE}}$
  is determined from a 
  high-frequency expansion in such a way, that the second derivatives of $\Delta
  G(\tau)$ vanish and the higher derivatives are reduced at the
  boundaries. Consequently, natural cubic splines are appropriate for
  $\Delta G$. The full FT to Matsubara
  frequencies is then obtained as $G(i\omega_n)=\Delta G(i\omega_n)+
  G_{\text{HFE}}(i\omega_n)$, where the last term is 
  computed directly via the Dyson equation from the model self-energy 
  $\Sigma_{\text{HFE}}(i\omega_n)$. The latter is constructed on the
  basis of the asymptotic expansion\cite{Knecht02,Oudovenko02}
  $\Sigma_{\gamma}(\omega) = \sum_{\beta\not=\gamma}U_{\beta\gamma}\expectv{n_{\beta}} + \Sigma_{\gamma}^{(2)}(\omega)+\mathcal{O}(\frac{1}{\omega^2})$ with
 \[
     \Sigma_{\gamma}^{(2)}(\omega)= \frac{1}{\omega} \sum\nolimits_{\alpha,\beta\not=\gamma} U_{\alpha\gamma}
       U_{\beta\gamma} (\expectv{n_{\alpha}n_{\beta}}
       - \expectv{n_{\alpha}}\expectv{n_{\beta}})\,.
  \]
  Here $\alpha, \beta$ and $\gamma$ are multi-indices, combining spin
  and orbital degrees of freedom: $\alpha\equiv(m,\sigma)$; the interaction
  matrix is defined by $U_{m\sigma,m\bar{\sigma}}=U$, 
  $U_{m\sigma,\bar{m}\sigma'}=U'-J_z\delta_{\sigma\sigma'}$.
  For the one-band model, where\cite{Potthoff97}
  $\Sigma_{\sigma}^{(2)}(\omega)=[U^2\expectv{n_{\bar{\sigma}}}(1-\expectv{n_{\bar{\sigma}}})]/\omega$,
  the exceptional accuracy of our DMFT-QMC method has already been
  established in comparisons to semi-analytic extrapolated
  perturbation theory for the insulating phase.\cite{Bluemer04}
  Since the algorithm reduces the discretization error to the inevitable
  Trotter error,  reliable QMC-data can be obtained
  already from a rather coarse discretization; the following results are
  based on QMC for $\dt=0.4$ unless indicated otherwise. 
  The DMFT cycle was typically iterated 20 times with $10^6$ sweeps 
  after convergency, using 1000 Matsubara frequencies. Full details of our 
  algorithm will be presented elsewhere.\cite{Knecht04}

  {\it Results --} In the following, we present QMC results 
  for the $J_z$-model with $J_z=U/4$, $U'=U/2$ and semi-elliptic densities
  of states with full bandwidths $W_1=2$, $W_2=4$, for the ``narrow'' and
  ``wide'' band, respectively.
 
  A traditional criterion for metal-insulator transitions is the
  quasiparticle weight $Z=[1-d \text{Re}\,\Sigma /d\omega|_{\omega=0}]^{-1}$ 
[for self-energy $\Sigma(\omega)$] which in the context of QMC simulations
  is estimated in a secant approximation as $Z\approx [1-
  \text{Im}\,\Sigma(i\pi T)/(\pi T)]^{-1}$. Evidently both definitions recover
  the limit $Z\to 1$ in the absence of interactions (when
  $\Sigma\equiv 0$); however, in the insulating regime the discrete
  approximation necessarily remains finite, whereas the true $Z$ vanishes
  exactly. As a consequence, metal-insulator transitions are expected
  to appear washed-out at finite temperatures. This is seen in the
  inset of \reff{Zratio}b for $T=1/32$:
  \begin{figure}
  \includegraphics[width=\columnwidth,clip=true]{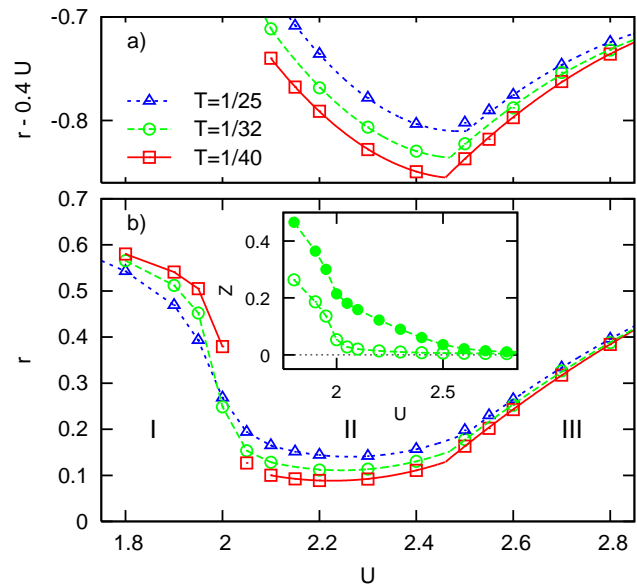}
  \caption{(Color online) b) Ratio 
    $r=Z_{\text{narrow}}/Z_{\text{wide}}$ of the discrete QMC
    estimates of the quasiparticle weights of both bands (shown in the
    inset) versus interaction $U$ for various temperatures $T$; lines
    for $U>2.1$ represent piecewise quadratic fits. a) Same data with
    linear offset.}
  \label{fig:Zratio}
  \end{figure}
  $Z$ drops only by about 60\% when the narrow band becomes insulating
  at $U\approx 2.0$. A second transition is not visible on this scale.
  The ratio $r\equiv Z_{\text{narrow}}/Z_{\text{wide}}$ shown in the
  main panel of \reff{Zratio}b is more illuminating since we can
  clearly distinguish three regions: in region I ($U\lesssim 2.0$),
  $r$ is of order unity with a sharp decrease near the boundary; in
  region II ($2.0\lesssim U\lesssim 2.5$), $r$ is nearly constant and
  of order 0.1, and in region III ($U\gtrsim 2.5$), $r$ increases with
  nearly constant slope (until it approaches a finite limit).  In
  order to analyze the boundary between II and III, we have performed
  piecewise quadratic fits to the QMC data for both regions separately
  (lines for $U>2.1$ in \reff{Zratio}b). The kinks at the boundary,
  barely visible on this scale, but resolved after subtracting
  a linear offset in \reff{Zratio}a, indicate a second transition 
  at least for $T\le 1/32$ (for which all QMC 
  data fall on the fit curves).

  The intrinsic ambiguities associated with the discrete estimate for
  $Z$ can be overcome in a full low-frequency analysis of the self-energy
  as shown in \reff{wImS}.
  \begin{figure}
  \includegraphics[width=\columnwidth,clip=true]{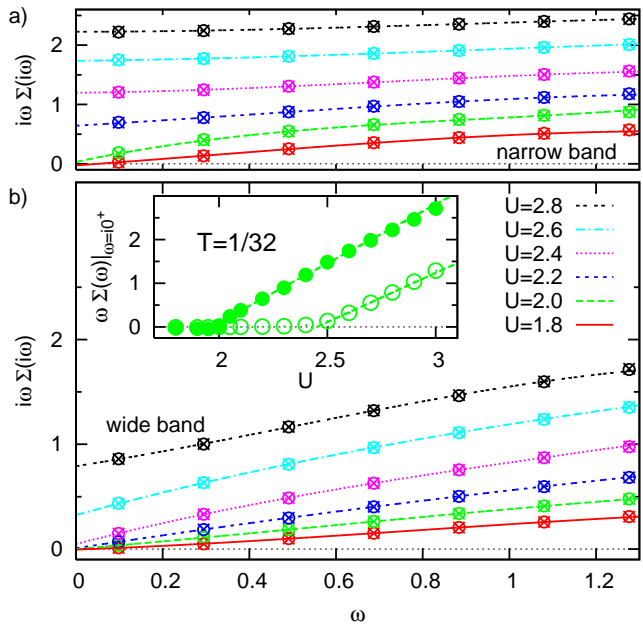}
  \caption{(Color online) Low-frequency analysis of the self-energy
    for $T=1/32$: QMC data (crosses: for $\dt=0.40$, circles: for
    $\dt=0.32$) for the product of frequency $\omega$ and self-energy
    $\Sigma$ at the Matsubara frequencies is extrapolated by cubic
    polynomials in $\omega$ for the narrow/wide band (upper/lower
    panel). Inset: extrapolated value for $\omega\to i0^+$.}
 \label{fig:wImS}
  \end{figure}
  Here, the data points represent QMC estimates of the product
  $\omega\Sigma(\omega)$ at the Matsubara frequencies $i\omega_n$;
  these products are real-valued since the self-energy is purely
  imaginary on the imaginary axis due to particle-hole symmetry. The
  lines, given by cubic polynomials in $\omega$, are expected to
  extrapolate to 0 at least linearly within a metallic phase (where $\Sigma$ is regular)
  and to a finite value within an insulating phase. By this criterion,
  the narrow band (cf.\ \reff{wImS}a) becomes insulating
  for $U\gtrsim 2.0$ while the wide band (cf.\ \reff{wImS}b) remains
  metallic up to $U=2.4$.  Note that the QMC results obtained from
  discretizations $\dt=0.4$ (crosses) and $\dt=0.32$ (circles) agree on the
  scale of the figure. The extrapolated product $\omega
  \Sigma(\omega)|_{\omega= i0^+}$, a measure of the singularity in the
  self-energy and roughly proportional to the expected gap, is shown
  at better resolution as a function of $U$ in the inset of
  \reff{wImS}b. We clearly see the presence of two distinct
  transitions for the narrow and wide band at $U=U_{c1}\approx 2.0$ and
  $U=U_{c2}\approx 2.5$, respectively. Corresponding results for
  higher and lower temperatures ($T=1/25$ and $T=1/40$, not shown) are
  barely distinguishable from \reff{wImS} except for $U\approx U_{c1}$.

  The spectral function $N(\omega)=-\text{Im}\, G(\omega)/\pi$, obtained
  from QMC by analytic continuation using the maximum-entropy method (MEM), 
  is depicted in \reff{spect}. 
  \begin{figure}
  \includegraphics[width=\columnwidth,clip=true]{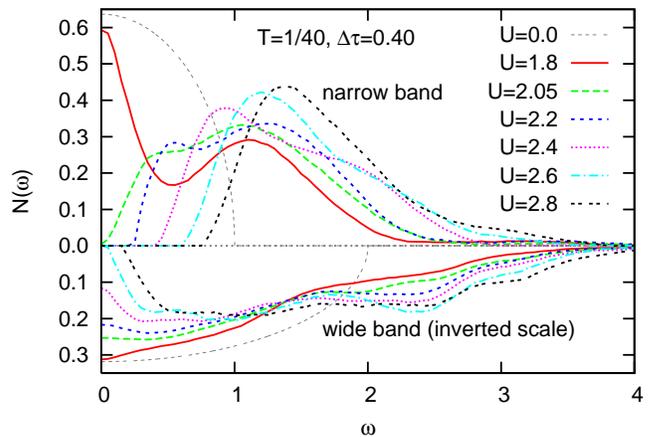}
  \caption{(Color online) Spectral function of narrow/wide band 
  (upper/lower panel) from QMC + MEM for $T=1/40$.}
  \label{fig:spect}
  \end{figure}
  \begin{figure}
  \includegraphics[width=\columnwidth,clip=true]{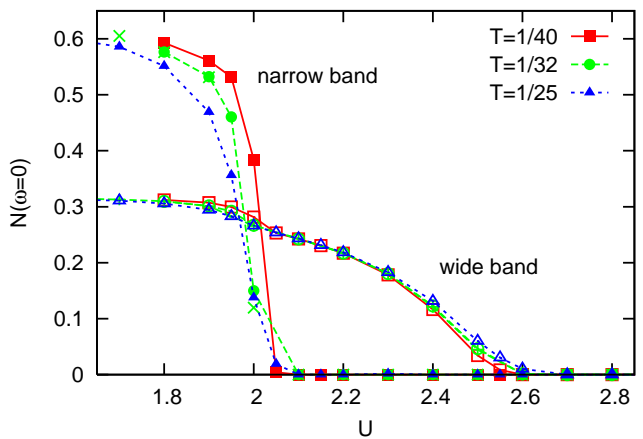}
  \caption{(Color online) QMC estimates for value $N(0)$ of spectral
  function at Fermi energy versus interaction. Full/empty symbols
  correspond to the narrow/wide band at $\dt=0.4$; crosses to results
  for $\dt=0.32$ at $T=1/32$.}
  \label{fig:N0}
  \end{figure}
  Evidently, $N(\omega\!=\!0)$ is accurately pinned at
  its noninteracting value up to $U=1.8$ for the wide band (lower
  panel) at $T=1/40$ and nearly pinned for the narrow band (upper
  panel). The narrow band develops a gap for $U>2.05$, where the wide
  band is still clearly metallic with a near-flat DOS at low
  frequencies. Only for $U=2.6$ a gap appears also for the wide band;
  note that the corresponding curve has a striking similarity to the
  narrow-band spectrum for $U=2.05$. Also noteworthy are the
  sharp band edges at low frequencies (where MEM is most reliable) and
  the rapid decay at high frequencies; apparently, the high
  quality of our data has kept artificial broadening to a minimum.
  The full dependence of $N(\omega\!=\!0)$ 
  on interaction and temperature is better resolved in \reff{N0},
  \begin{figure}
  \includegraphics[width=\columnwidth,clip=true]{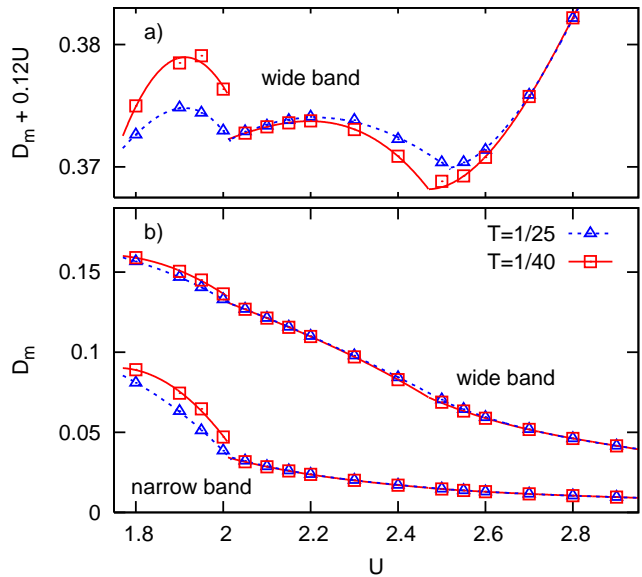}
  \caption{(Color online) Intraorbital double occupancy 
   $D_m=\langle n_{m\uparrow} n_{m\downarrow}\rangle$. Lines indicate 
    piecewise cubic fits. Upper
    panel: a linear offset exposes the features of the wide band.}
  \label{fig:D}
  \end{figure}
  which clearly exposes the orbital-selective Mott transition: at 
  $U_{c1}$ a sudden decay to 0 is observed only for the narrow band,
  while the wide-band value is reduced only by some $20\%$. Evidently,
  the second band becomes insulating only for $U_{c2}\approx 2.5$. Both
  transitions become sharper at lower temperatures. A discretization
  dependence is visible for $T=1/32$ only at $U=2.0$.
  
  Finally, we discuss the
  intraorbital double occupancy $D_m=\langle n_{m\uparrow}
  n_{m\downarrow}\rangle$. As seen in \reff{D}b,
  $D_{\text{wide}}$ barely shows any features near $U_{c1}$, while
  $D_{\text{narrow}}$ is reduced by about 50\% from $U\!=\!1.9$ to
  $U\!=\!2.1$.  We conclude that the wide orbital remains itinerant
  at a point where the narrow orbital is already localized. 
  A second transition of the wide band is clearly seen only after
  adding a suitable linear term in \reff{D}a. Here, both
  transitions appear as regions of enhanced temperature
  dependence and as kinks. Note that up to the
  discontinuity at $U_{c1}$ for $T\!=\!1/40$ and on a broader scale the
  kink is even more pronounced at $U_{c2}$ than at $U_{c1}$. 
  Once again we find {\em two} phase transitions, each of which can be
  associated with a Mott transition of one orbital. Signatures in
  observables associated with the other orbital reflect
  the fact that phase transitions usually leave traces
  in every observable.

  {\it Discussion and Outlook --} 
  We considered a two-band Hubbard model with distinct band widths
  $W_2=2 W_1$ and interaction parameters $U,U'\!=\!U/2,J_z\!=\!U/4$.
  Using high-precision QMC calculations, in which the correct
  high-frequency behavior of the self-energy is carefully taken into
  account, we showed that this $J_z$-model contains \emph{two}
  successive metal-insulator transitions and can hence be considered
  as a minimal model for the OSMTs, observed experimentally in 
  Ca$_{2-x}$Sr$_{x}$RuO$_4$. These transitions are particularly
  clearly revealed by the low-frequency behavior of the self-energy, but
  also visible in the spectral functions,
  quasiparticle weights, and intraorbital double occupancies.
  Our high-precision data correct earlier QMC results
  \cite{Liebsch03a, Liebsch03b, Liebsch04} by Liebsch. They also
  contradict early ED results \cite{Koga04b} by Koga {\em et al.}, who may
  have missed the relatively narrow OSMT region due to ED's finite
  energy resolution. On the other hand, our results are in qualitative
  agreement with de' Medici {\em et al.}'s recent slave-spin mean-field
  phase diagram for the $J_z$-model,\cite{deMedici05} which also
  contains an orbital-selective Mott phase, albeit at slightly larger
  $U$-values (at $T=0$). 
 
  Our results imply in particular that \emph{isotropy} of the Hund
  exchange is \emph{not} a prerequisite for the existence of OSMTs, as
  was suggested by Koga {\em et al.}\cite{Koga04b} on the basis of their ED
  study. In future work it will be of interest to investigate the
  effect of other terms in the Hamiltonian, such as hybridization terms
   of various symmetry;
  first results in this direction have already been reported in 
  Ref.\ \onlinecite{Koga05}. It would also be of interest to extend the
  calculations reported here to lower temperatures, to the more
  realistic 3-band case and to magnetic phases.

  {\it Acknowledgements --}
  We thank K.~Held, E.~Jeckelmann, A.~Liebsch, and K.~Ueda for stimulating
  discussions and acknowledge support by the NIC J\"ulich and by the DFG
  (Forschergruppe 559, Bl775/1).


\begin{thebibliography}{16}
\bibitem{fn:IPT} Our reanalysis of the IPT data of Ref.\ \onlinecite{Liebsch04} revealed
a sharp second transition at $U\approx 3.6-3.7$ for low $T\le 0.02$.
\end{thebibliography}
 \end{document}